\documentclass[prb,showpacs,twocolumn,aps,superscriptaddress,a4paper]{revtex4-1}
\usepackage{dcolumn,amssymb,amsmath,amsfonts,graphicx,latexsym,color,braket}

\definecolor{myblue}{rgb}{0,0,0.75}

\begin{document}

\title{Poincar\'{e} crystal on the one-dimensional lattice}

\author{Pei Wang}
\email{wangpei@zjnu.cn}
\affiliation{Department of Physics, Zhejiang Normal University, Jinhua 321004, China}

\date{\today}

\begin{abstract}
In this paper, we develop the quantum theory of particles that has discrete Poincar\'{e} symmetry
on the one-dimensional Bravais lattice. We review the recently discovered discrete Lorentz symmetry,
which is the unique Lorentz symmetry that coexists with the discrete space translational symmetry
on a Bravais lattice. The discrete Lorentz transformations and spacetime translations form the discrete Poincar\'{e} group,
which are represented by unitary operators in a quantum theory.
We find the conditions for the existence of representation, which are expressed as the congruence relation
between quasi-momentum and quasi-energy.
We then build the Lorentz-invariant many-body theory of indistinguishable particles by expressing both the unitary operators
and Floquet Hamiltonians in terms of the field operators. Some typical Hamiltonians include the long-range hopping which
fluctuates as the distance between sites increases. We calculate the Green's functions of the
lattice theory. The spacetime points where the Green's function is nonzero display a lattice structure. During the
propagation, the particles stay localized on a single or a few sites to preserve the Lorentz symmetry.
\end{abstract}

\maketitle

\section{Introduction}

The realization of intriguing lattice models, for examples, the Bose-Hubbard~\cite{Greiner} and Haldane~\cite{Jotzu} models,
with ultracold atoms in optical lattices~\cite{Lewenstein07} encourages
researchers to continuously explore new quantum lattice theories with exotic properties. The
spacetime translation, rotation, space inversion and time reversal symmetries all play
important roles in the construction of a lattice theory. The discrete space-translational symmetry
causes the continuous energy of a free particle broken into a series of Bloch bands~\cite{Girvin}.
The rotation and parity symmetries constrain the spectrum of particles.
The discrete time-translational symmetry in periodically-driven systems requires that the
eigenstate and energy be replaced by Floquet state and quasi-energy, respectively,
and the latter was employed to classify the Floquet topological phases of matter in recent years~\cite{Kitagawa}.
The time reversal symmetry leads to the Kramers' degeneracy which is the key for understanding
the topological insulators~\cite{Hasan}. But the Lorentz symmetry has always been ignored in
condensed matter physics. A question naturally arises as to whether there exists a quantum lattice model
that has the Lorentz symmetry.

The Lorentz symmetry is essentially important in quantum field theory which is indeed the result of efforts in reconciling
quantum mechanics with Lorentz symmetry~\cite{Weinberg}.
Wigner's theorem paved the way for representing physical symmetry transformations by
the unitary or antiunitary operators in the Hilbert space~\cite{Wigner31}. The quantum field theories keep
invariant under arbitrary spacetime translations and Lorentz transformations which form
a Lie group - the Poincar\'{e} group. The existence of elementary particles with different spin or helicity
can be directly deduced from the fact that the Poincar\'{e} group has different irreducible representations.
On the other hand, in the study of crystalline materials or optical lattices which have the
discrete space-translational symmetry like a Bravais lattice,
the low-energy effective field theory does not conform to the Lorentz invariance, because
the continuous Lorentz symmetry does not survive on a Bravais lattice.

One way of resurrecting the Lorentz symmetry on a lattice is by employing a random lattice
with the spacetime points sprinkled following a Poisson process~\cite{Bombelli}. Such a lattice
was introduced in the causal set approach to quantum gravity, which preserves Lorentz invariance on average
(see Ref.~\onlinecite{Surya} for a recent review). For strictly-preserved symmetries,
the discrete Lorentz symmetry was found recently~\cite{Wang18,Wang20}, which coexists with the discrete
space-translational symmetry. A Bravais lattice of constant $a$
keeps invariant under a cyclic group of Lorentz transformations in spite of length contraction.
From classical point of view, this is possible if the lattice is shaken periodically.
In one frame of reference, $a$ is the spatial distance between two nearest-neighbor sites
that are oscillating in the same phase. In the second frame of reference that is moving
relative to the first one, these two sites have different phases due to the relativity of simultaneity so that they are not
"nearest neighbors" any more. But two sites at original distance $na$ have the same phase now,
while their distance in the second reference frame becomes $a$ due to contraction.
The periodicity along the time direction requires that the Bravais lattice be extended into a spacetime lattice.
In the language of group, the discrete Lorentz transformations together with the discrete translations on the
spacetime lattice form a group - the discrete Poincar\'{e} group which is a subgroup of the continuous Poincar\'{e} group.
If a physical theory keeps the same form under arbitrary coordinate transformation in the discrete Poincar\'{e}
group, we say that the theory has the discrete Poincar\'{e} symmetry.

In this paper, we develop a general approach to construct the quantum theories that have the discrete Poincar\'{e} symmetry
on a one-dimensional Bravais lattice of length $N$. We start from the representations of discrete Poincar\'{e}
group and utilize the finite-dimensional property of the single-particle Hilbert space. Our approach is
exhaustive. The Lorentz invariance is translated into a self-consistent equation of quasi-energy and
quasi-momentum, which is further changed into a congruence relation modulo $N$ between integer variables.
This congruence equation has solutions for some but not all $N$. As examples, we enumerate
the $N$ ($N<100$) for which the solutions exist, and study the corresponding dispersion relations of a single particle.
We then turn into the many-body theory of indistinguishable bosons or fermions, obtaining
the expression of discrete Lorentz transformations and spacetime translations in terms of the
field operators. It is appropriate to call such a many-body model a Poincar\'{e} crystal.
For the convenience of designing an experiment, we construct the Floquet Hamiltonians
of the Poincar\'{e} crystals, which might possibly be realized with ultracold atoms in optical lattices.
The properties of a Poincar\'{e} crystal are studied. We calculate the Green's functions, which
exactly reflect the underlying Lorentz symmetry.

The paper is organized as follows. In Sec.~\ref{sec:review}, we review the definitions of
discrete Lorentz transformations and discrete Poincar\'{e} group. Sec.~\ref{sec:rep} introduces
the representations of discrete Poincar\'{e} group which determine how a single-particle
quantum state transforms under the unitary operators in the group. In Sec.~\ref{sec:singleH}, we construct the orthonormal
basis of the single-particle Hilbert space. In this process, the conditions for quasi-energy and quasi-momentum
are found. The many-body theory of Poincar\'{e} crystal is developed
in Sec.~\ref{sec:Hamiltonian} in which the field operators are defined, the symmetry transformations are
expressed in terms of them, and the Floquet Hamiltonians are presented. Sec.~\ref{sec:green} discusses
the Green's functions of a Poincar\'{e} crystal. Finally, Sec.~\ref{sec:summary} summarizes our results.

\section{Discrete Poincar\'{e} group \label{sec:review}}

The discrete Poincar\'{e} group is a group of coordinate transformations introduced in
Ref.~[\onlinecite{Wang18}] and~[\onlinecite{Wang20}]. In this section, we review its definition and
properties.

We consider the 1+1-dimensional spacetime, and use $x^\mu$ with $\mu=0,1$
to denote the time and space coordinates, respectively. Sometimes, we use
$t$ and $x$ instead of $x^0$ and $x^1$. The spacetime metric $\eta_{\mu\nu}$
is diagonal with elements $\eta_{00}=-1$ and $\eta_{11}=1$. A Lorentz transformation is a rotation in the
spacetime. It is represented by a 2-by-2 matrix that acts on the vector $\left(t,x\right)^{\text{T}}$, reading
\begin{equation}\label{eq:Lorentz}
L = \left(\begin{array}{cc}
\gamma & \text{sign}(v) \sqrt{\gamma^2-1}/c \\
\text{sign}(v) \sqrt{\gamma^2-1} \ c & \gamma
\end{array}\right),
\end{equation} 
where $\gamma=1/\sqrt{1-v^2/c^2}\geq 1$ is dimensionless and $v$ is the velocity
of one reference frame with respect to the other. $c$ is the speed of light
in special theory of relativity. In this paper, we treat
$c$ as a speed of constant used to define the Lorentz transformation, but
the value of $c$ can be chosen arbitrarily. The Lorentz transformations with $-c<v<c$
form a continuous group.

The discrete Lorentz group is a subgroup of the continuous Lorentz group. Its elements are
\begin{equation}\label{eq:disLorentz}
L(j) = \left(\begin{array}{cc}
\gamma(j) & \text{sign}(j) \sqrt{\gamma(j)^2-1}/c \\
\text{sign}(j) \sqrt{\gamma(j)^2-1} \ c & \gamma(j)
\end{array}\right),
\end{equation} 
where
\begin{equation}\label{eq:gammaj}
\begin{split}
\gamma(j)= \frac{1}{2} \left(\gamma(1)  +\sqrt{\gamma(1)^2-1}\right)^j 
+ \frac{1}{2} \left(\gamma(1)-\sqrt{\gamma(1)^2-1}\right)^j
\end{split}
\end{equation}
and $j=0, \pm 1, \pm 2,\cdots$ is an integer. $\gamma(1)>1$ is an integer
or half-integer, dubbed the generator. A discrete Lorentz group is uniquely determined by $\gamma(1)$.
Interestingly, $\gamma(j)$ at arbitrary $j$
can be obtained by using the iterative relation $\gamma(j+1) = 2\gamma(1)\gamma(j)-\gamma(j-1)$,
given $\gamma(0)=1$ and $\gamma(1)$. In this paper, we consider an integer generator
($\gamma(1)=2, 3, \cdots$), hence $\gamma(j)$ is an integer. We define
$\zeta(j)=\text{sign}(j)\sqrt{\gamma(j)^2-1}/\sqrt{\gamma(1)^2-1}$, which satisfies
a similar relation $\zeta(j+1)=2\gamma(1)\zeta(j)-\zeta(j-1)$ with $\zeta(0)=0$
and $\zeta(1)=1$. One can prove that $\zeta(j)$ is an integer.
Some useful formulas about $\gamma(j)$ and $\zeta(j)$ are listed in Appendix~\ref{app:gamma}.
Now we reexpress the discrete Lorentz transformation as
\begin{equation}\label{eq:disLorentzzeta}
L(j) = \left(\begin{array}{cc}
\gamma(j) & \zeta(j) \sqrt{\gamma(1)^2-1} /c \\
\zeta(j) \sqrt{\gamma(1)^2-1} \ c & \gamma(j)
\end{array}\right).
\end{equation} 
An important property of $L(j)$ is that it is equal to $L(1)$ to the power $j$.
As a consequence, we have $L(j+j')=L(j)L(j')$. The discrete Lorentz group is a cyclic group.

A general coordinate transformation is a combination of Lorentz transformation and
spacetime translation, which is expressed as
\begin{equation}\label{eq:Poincare}
 \left(\begin{array}{c}t' \\ x' \end{array}\right)=L\left(\begin{array}{c}t \\ x \end{array}\right)
 + \left(\begin{array}{c}\Delta t \\ \Delta x \end{array}\right),
\end{equation}
where $\left(\Delta t,\Delta x\right)^\text{T}$ is the translation vector. We use $P$ to denote
such a transformation. $P$ maps $\left(t,x\right)^{\text{T}}$ into $\left(t',x'\right)^{\text{T}}$.
This fact can be used to derive the rule of multiplication for $P$. All the $P$s form a continuous group -
the so-called Poincar\'{e} group, which is well known in special relativity and quantum field theory.

\begin{figure}[tbp]
\includegraphics[width=0.9\linewidth]{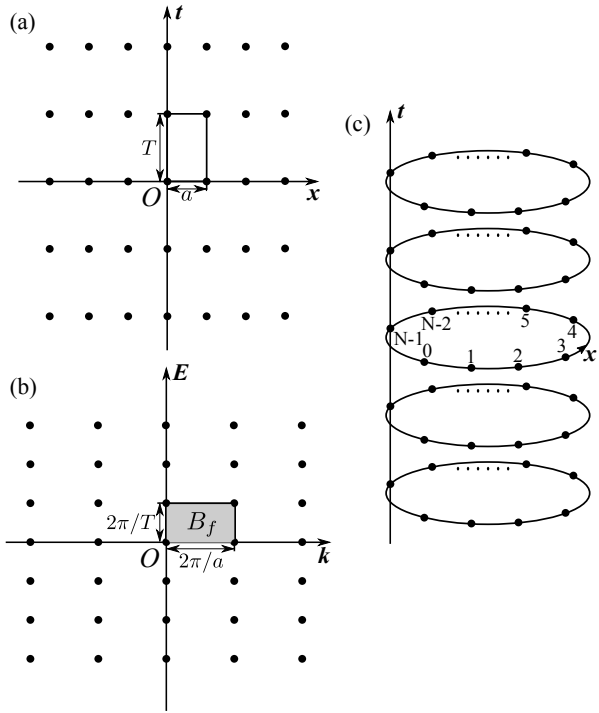}
\caption{The schematic diagram of (a) a spacetime Bravais lattice $\mathcal{Y}$, (b)
the corresponding reciprocal lattice $\mathcal{R}$ and (c) a lattice of finite size with
periodic boundary conditions along the spatial direction.}\label{fig:lattice}
\end{figure}

The discrete Poincar\'{e} group, denoted by $\mathcal{P}$, is a subgroup of the continuous Poincar\'{e} group.
$\mathcal{P}$ includes only discrete Lorentz transformations and discrete spacetime translations.
We use $a$ to denote the period of space translation, and define
$T=\sqrt{\gamma(1)^2-1} \ a/c$ to be the period of time translation. The primitive
vectors ${\bf e} = (T,0)^\text{T}$ and ${\bf f}=(0,a)^\text{T}$ generate
a 1+1-dimensional rectangular lattice, dubbed $\mathcal{Y}$
(see fig.~\ref{fig:lattice}(a)). The elements of $\mathcal{P}$ are written
as $P(j,m,n)$ with $j,m,n=0, \pm 1, \pm 2, \cdots$ being integers. $P(j,m,n)$ denotes
a Lorentz transformation $L(j)$ followed by a translation of vector $m {\bf e} + n{\bf f}$.
For examples, $P(j,0,0)$ is indeed $L(j)$, and $P(0,m,0)$ and $P(0,0,n)$
are pure temporal and spatial translations, respectively.
One can prove that the lattice $\mathcal{Y} $ is invariant under $L(j)$. This guarantees
that $\left\{ P(j,m,n)\right\}$ is a group. By using the definition of $P(j,m,n)$ and
the properties of $L(j)$ and by studying the transformation of $m {\bf e} + n{\bf f}$
under $L(j)$, we obtain the group's multiplication laws, which are listed below:
\begin{equation}\label{eq:Pmultilaw}
\begin{split}
& P(j,m,n)= P(0,m,n) P(j,0,0), \\
& P(j,0,0) = P^j (1,0,0) , \\
& P(0,m,n)= P^m(0,1,0) P^n(0,0,1), \\
& \left[ P(0,1,0), P(0,0,1) \right] = 0 , \\
& P(j,0,0) P(0,m,n) P^{-1}(j,0,0) = P(0,m',n') \\
& with \ \left( \begin{array}{c} m' \\ n' \end{array}\right) =  \left(
\begin{array}{cc}  \gamma(j) & \zeta(j) \\ \zeta(j) \left(
\gamma(1)^2-1\right) & \gamma(j) \end{array} \right) \left( \begin{array}{c}
m \\ n \end{array} \right),
\end{split}
\end{equation}
where $P^{-1}$ denotes the inverse of $P$ and $\left[ P, P' \right]$ denotes
the commutator of $P $ and $P'$. It is easy to verify that $m'$ and $n'$ are
integers given an integer pair $(m,n)$, since $\zeta(j)$ and $\gamma(j)$
are both integers. Eq.~\eqref{eq:Pmultilaw} can be viewed as the defining properties
of $\mathcal{P}$.

A few comments are necessary. (i) $P(1,0,0)$, $P(0,1,0)$ and $P(0,0,1)$ are the three generators of $\mathcal{P}$.
The other elements of $\mathcal{P}$ can be expressed as the product and power of them.
(ii) $\mathcal{P}$ is uniquely determined by $\gamma(1)$, $c$ and $a$.
Two discrete Poincar\'{e} groups with the same $\gamma(1)$ are isomorphic to each other.
(iii) The discrete Poincar\'{e} groups are the only groups
that contain at least one non-identity Lorentz transformation at the same time whose spatial
translations are discrete. In other words, if we want to keep some Lorentz
symmetry on a lattice model, its symmetry group must be a discrete Poincar\'{e} group.

\section{Representation of discrete Poincar\'{e} group \label{sec:rep}}

In quantum mechanics, a physical state is a vector in the Hilbert space $\mathcal{H}$.
A symmetry transformation is a unitary operator mapping $\mathcal{H}$ to itself.
To construct a quantum theory with discrete Poincar\'{e} symmetry is equivalent
to find a unitary representation of the group $\mathcal{P}$. We set $\hbar= 1$ throughout the paper.

If $P(j,m,n) \in \mathcal{P}$ is the transformation from the coordinate system of observer
$K$ to that of observer $K'$, we use $\hat U(j,m,n)$ to denote the unitary transformation mapping
the quantum states seen by $K$ to those seen by $K'$.
$\hat U(j,m,n)$ obeys the multiplication laws as same as Eq.~\eqref{eq:Pmultilaw}. To save time, we do not
write them down again. Let us consider the classification of single-particle states
according to their transformation under $\hat U(j,m,n)$. Since $\hat U(0,m,n)$ of different $(m,n)$
all commute with each other, they have common eigenstate, denoted as $\ket{p}$.
According to $\hat U(0,m,n)=\hat U^m(0,1,0) \hat U^n(0,0,1)$, the eigenvalue
of $\hat U(0,m,n)$ for arbitrary $(m,n)$ is determined by the eigenvalues of $\hat U (0,1,0)$ and $\hat U(0,0,1)$.
The eigenvalue of a unitary operator must be a complex number of unit magnitude. Without loss
of generality, we assume the eigenvalues of $\hat U (0,1,0)$ and $\hat U(0,0,1)$
to be $e^{-i \eta_{\mu\nu} e^\mu p^\nu}$ and $e^{-i \eta_{\mu\nu} f^\mu p^\nu}$, respectively, where
$e^\mu$ and $f^\mu$ are the components of the translation vectors ${\bf e}$ and ${\bf f}$, respectively,
and the summation convention has been applied. This assumption makes it clear that
$p$ should be a 1+1-dimensional vector with $p^0$ and $p^1$ being its components.
Sometimes, we also use $E= p^0$ and $k= p^1$ to denote the components of $p$.
Now we have
\begin{equation}\label{eq:eigenvalueUmn}
\begin{split}
\hat U(0,m,n) \ket{p} = & e^{-i \eta_{\mu\nu} \left( m {e}+n {f}\right)^\mu p^\nu} \ket{p} \\
= & e^{-i n a k +i m TE} \ket{p}.
\end{split}
\end{equation}
In Eq.~\eqref{eq:eigenvalueUmn}, $\left( m {\bf e}+n {\bf f}\right)$ is recognized as the translation
vector of $P(0,m,n)$. Taking a comparison with quantum field theory, one may naively
identify $E$ as the "energy" and $k$ as the "momentum". We emphasize that the energy or momentum
may not be well-defined if there does not exist a continuous spacetime-translational symmetry.
In the presence of discrete translational symmetry, it is more appropriate to call $E$
the "quasi-energy" and $k$ the "quasi-momentum".

To decide how $\ket{p}$ transforms under $\hat U(j,0,0)$, we utilize the last multiplication
law in Eq.~\eqref{eq:Pmultilaw} by rewriting it as
\begin{equation}\label{eq:Umultilaw}
\hat U (j,0,0) \hat U(0,m,n) = \hat U (0,m',n') \hat U(j,0,0) .
\end{equation}
Acting both sides on $\ket{p}$, we find that $\hat U(j,0,0) \ket{p}$ is an eigenvector
of $\hat U (0,m',n')$ with the eigenvalue $e^{-i \eta_{\mu\nu} \left( m {e}+n {f}\right)^\mu p^\nu}$.
Without considering the internal degrees of freedom (spin or helicity), we can simply assume
\begin{equation}\label{eq:Ujonp}
\hat U(j,0,0) \ket{p} = \ket{p'} ,
\end{equation}
where $p'$ satisfies 
\begin{equation}\label{eq:pandpp}
e^{-i \eta_{\mu\nu} \left( m {e}+n {f}\right)^\mu p^\nu}=
e^{-i \eta_{\mu\nu} \left( m' {e}+n' {f}\right)^\mu p'^\nu}.
\end{equation}

Before giving the solution of Eq.~\eqref{eq:pandpp}, we define the reciprocal lattice of $\mathcal{Y}$,
denoted by $\mathcal{R}$ whose two primitive vectors (denoted by ${\bf g}$ and ${\bf h}$)
satisfy $-\eta_{\mu\nu} e^\mu g^\nu = \eta_{\mu\nu} f^\mu h^\nu = 2\pi$ and
$\eta_{\mu\nu} e^\mu h^\nu = \eta_{\mu\nu} f^\mu g^\nu = 0$. ${\bf g}$ and ${\bf h}$
are found to be ${\bf g}= (2\pi/T,0)^\text{T}$ and ${\bf h}=(0,2\pi/a)^\text{T}$, respectively.
Obviously, $\mathcal{R}$ is a rectangular lattice (see fig.~\ref{fig:lattice}(b)). And for arbitrary lattice vectors
$r\in \mathcal{R}$ and $y\in \mathcal{Y}$, $\eta_{\mu\nu}r^\mu y^\nu$
must be integer times of $2\pi$. $\mathcal{Y}$ is a lattice located on the $t$-$x$ plane (real spacetime), while
$\mathcal{R}$ is a lattice located on the $E$-$k$ plane.

$\ket{p}= \ket{E,k}$ transforms into $\ket{p'}= \ket{E',k'}$ under the unitary transformation $\hat U(j,0,0)$.
The solution of Eq.~\eqref{eq:pandpp} can be expressed as (see Appendix~\ref{app:Ektran} for the derivation)
\begin{equation}\label{eq:Ektransform}
\left( \begin{array}{c} E' \\ k' \end{array} \right) \stackrel{\mathcal{R}}{=} L(j) 
\left( \begin{array}{c} E \\ k \end{array} \right),
\end{equation}
where $\stackrel{\mathcal{R}}{=} $ denotes an equivalence relation. Two vectors on the
$E$-$k$ plane are equivalent to each other if and only if their difference is a lattice vector
of $\mathcal{R}$ (i.e. $m {\bf g} + n {\bf h}$ with $m$ and $n$ being integers).
Eq.~\eqref{eq:Ektransform} tells us how the quasi-energy-momentum transforms under a Lorentz transformation.

We then divide the $E$-$k$ plane into a series of Brillouin zones (unit cells of the reciprocal lattice).
For convenience, we work in the Brillouin zone denoted as $B_f$ that is
the rectangle with edges $0\leq E<2\pi/T$ and $0\leq k<2\pi/a$ (see fig.~\ref{fig:lattice}(b)).
The quasi-energy-momentum $(E,k)$ must be in $B_f$.
$B_f$ is the generalization of a 1D crystal's Brillouin zone into 1+1 dimensions.
A crystal has only the space-translational symmetry, but here we consider both the space
and time translation symmetries, therefore, $B_f$ is 1+1-dimensional.

Sometimes, the formulas are simplified by introducing the dimensionless quasi-energy and
quasi-momentum, defined as $\bar E=\displaystyle\frac{E}{(2\pi/T)}$ and $\bar k=
\displaystyle \frac{k}{(2\pi/a)}$, respectively. Correspondingly, the dimensionless reciprocal lattice
is denoted as $\bar{ \mathcal{R}}$ whose primitive vectors are simply $(1,0)^\text{T}$ and $(0,1)^\text{T}$.
And the Brillouin zone becomes a unit square with $0\leq \bar E<1$ and $0\leq \bar k <1$.
On the $\bar E$-$\bar k$ plane, Eq.~\eqref{eq:Ektransform} can be reexpressed as
\begin{equation}\label{eq:Ebarkbartrans}
\begin{split}
\left( \begin{array}{c} \bar E' \\ \bar k' \end{array} \right) \stackrel{\mathcal{\bar R}}{=} 
\left( \begin{array}{cc} \gamma(j) & \zeta(j) \left(\gamma(1)^2-1\right) \\
\zeta(j) & \gamma(j) \end{array}\right)
\left( \begin{array}{c} \bar E \\ \bar k \end{array} \right),
\end{split}
\end{equation}
where $( \bar E,\bar k)$ and $(\bar E', \bar k')$ are both in the Brillouin zone.
An important property of Eq.~\eqref{eq:Ebarkbartrans} is that the matrix elements
at its right-hand side are all integers.

\section{Single-particle Hilbert space \label{sec:singleH}}

\subsection{How to construct the Hilbert space}

The eigenvectors of a unitary operator with different eigenvalues
are orthogonal to each other. $\ket{p}$ is the eigenvector of $\hat U(0,m,n)$.
For $p,p'\in B_f$, we then have $\braket{p|p'} = \delta_{p,p'}$.
Suppose the dimension of the single-particle Hilbert space is finite, denoted as $N$.
We then choose $N$ different $\ket{p}$ with $p\in B_f$ to form an orthonormal basis of the single-particle
Hilbert space.

There exist some conditions for the set $\left\{ \ket{p}\right\}$ being a basis.
To obtain these conditions, we study the single-particle basis in real spacetime. Since the single-particle Hilbert space has $N$
dimensions, our model must contain $N$ spatially-different sites on which the particle can be located.
Without loss of generality, we number these sites from $0$ to $N-1$ whose positions are correspondingly
$x=0, a, 2a, \cdots, (N-1)a$. To maintain the space-translational symmetry, we adopt the periodic boundary
condition along the $x$-axis, i.e., acting $P(0,0,1)$ on the site at $x=(N-1)a$ results in the site at $x=0$.
In other words, we treat two spacetime sites $(mT, na)$ and $(mT, n' a)$ in fig.~\ref{fig:lattice}(a) as being equivalent to each other
if $n\equiv n' \ \left(\text{mod} \ N\right)$. The quotient set of $\mathcal{Y}$ with respect to this equivalence relation
is topologically equivalent to a cylinder (see fig.~\ref{fig:lattice}(c)), which contains the sites $\left(mT, na\right)$
with $-\infty < m<\infty$ but $0 \leq n \leq N-1$.

We use $\ket{m,n}$ to denote a physical state made by creating a particle at
the position $x=na$ and the time $t=mT$. Notice that we have chosen the Heisenberg picture here (the state
does not change with time). Different observers see different state-vectors.
Suppose $\ket{m,n}$ is the state-vector seen by an observer $K$.
The coordinate transformation from $K$ to another observer $K'$ is $t'=t+m'T$ and $x'=x+n'a$.
This transformation is a translation of vector $m' {\bf e}+n' {\bf f}$, therefore,
the state-vector seen by $K'$ should be $\hat U(0,m',n')\ket{m,n}$. On the other hand, the observer $K'$ sees
that this particle is created at the position $x'=(n+n')a$ and the time $t'=(m+m')T$, which indicates
\begin{equation}\label{eq:realstatetranslation}
\ket{m+m',n+n'} = \hat U(0,m',n') \ket{m,n}.
\end{equation}
Eq.~\eqref{eq:realstatetranslation} defines how a basis vector in real spacetime transforms under
the translations. As a single-particle state, $\ket{m,n}$ can be decomposed into a linear combination
of basis vectors in the quasi-energy-momentum space. Without loss of generality, we
assume $\ket{0,0}=\sum_p \mathcal{N}_p \ket{p}$ where $\mathcal{N}_p$ is an undetermined coefficient. By using both
Eq.~\eqref{eq:eigenvalueUmn} and~\eqref{eq:realstatetranslation}, we obtain
\begin{equation}\label{eq:realdecomp}
\ket{m,n} = \sum_p \mathcal{N}_p e^{ -i \eta_{\mu\nu} \left(m e +n f \right)^\mu p^\nu} \ket{p}.
\end{equation}

For a given $m$ (e.g. $m=0$), $ \ket{m,n}$ with $n=0,1,\cdots N-1$ represents a set
of states in which the particle is located on different sites. They should form an orthonormal basis
of the single-particle Hilbert space. The orthogonality and completeness require $\braket{m,n|m,n'}
=\delta_{n,n'}$ and $\sum_{n=0}^{N-1} \ket{m,n}\bra{m,n} = 1$, respectively. By substituting Eq.~\eqref{eq:realdecomp}
in and using the relation $\sum_p \ket{p}\bra{p}=1$ ($\left\{\ket{p}\right\}$ is an orthonormal basis), we obtain
\begin{equation}\label{eq:orthonormal}
\begin{split}
& \sum_p \left|\mathcal{N}_p \right|^2 e^{i a p^1 \left(n-n'\right)} = \delta_{n,n'}, \\
& \sum_{n} \mathcal{N}_p \mathcal{N}_{p'}^* e^{i \eta_{\mu\nu} \left( me+nf\right)^\mu \left( p'-p\right)^\nu}
 = \delta_{p,p'}.
\end{split}
\end{equation}
Eq.~\eqref{eq:orthonormal} is the orthonormal condition for $\left\{ \ket{p}\right\}$ being a basis. Neglecting an unimportant phase,
we derive $\mathcal{N}_p = 1/\sqrt{N}$ from Eq.~\eqref{eq:orthonormal}. More important,
Eq.~\eqref{eq:orthonormal} puts a constraint on our choice of
$p^1$ or $k$. The $N$ quasi-momenta must be evenly-spaced, being expressed as
$k= \delta, \delta+\displaystyle\frac{2\pi}{Na}, \delta+2\displaystyle\frac{2\pi}{Na}, \cdots, \delta+\left(N-1\right)
\displaystyle\frac{2\pi}{Na}$ where $0\leq \delta< \displaystyle\frac{2\pi}{Na}$ is a tunable shift.
In this paper, we focus on the case $\delta=0$ in which the particle is allowed to
have zero quasi-momentum. It is worth mentioning that there exist theories with nonzero $\delta $ (e.g. $\delta=\pi/ Na$).

Since the $N$ (dimensionless) quasi-momenta are all different from each other,
we use $\ket{\bar{E}_{\bar k}, \bar{k} }$ with $\bar k=0, 1/N, \cdots, \left(N-1\right)/N$
to denote the basis vectors in the dimensionless quasi-energy-momentum space.
The orthonormal condition does not put any constraint on the choice of $\bar{E}_{\bar k}$ (the dispersion relation).
But there exists the other condition that constrains the dispersion relation.

For a quantum theory that has discrete Poincar\'{e} symmetry, the Hilbert space must
keep invariant under the transformations $\hat U(j,m,n)$. In other words, $\hat U(j,m,n)$
transforms each basis vector into another vector in the same Hilbert space.
The single-particle Hilbert space
is generated by a set of $\ket{p}$. According to Eq.~\eqref{eq:eigenvalueUmn},
$\ket{p}$ transforms into itself with an additional phase under the translations $\hat U(0,m,n)$,
hence, the Hilbert space naturally keeps invariant under $\hat U(0,m,n)$.
But $\ket{p}$ transforms into $\ket{p'}$ under the Lorentz transformation $\hat U(j,0,0)$.
The set $\left\{ \ket{p}\right\}$ must keep invariant under $\hat U(j,0,0)$ if it is a basis.
Since $\hat U(j,0,0)=\hat U^j(1,0,0)$, it is sufficient to require $\left\{ \ket{p}\right\}$ being invariant under $\hat U(1,0,0)$.
This condition is called the invariant condition, which is highly nontrivial.
In a quantum field theory with continuous Poincar\'{e} symmetry,
it is impossible to construct a finite-dimensional single-particle Hilbert space, because
a nonzero $p^1$ can be transformed into arbitrary momentum under
suitable Lorentz transformation and there exist infinite possible momenta ($p^1\in (-\infty, \infty)$).
But in a theory with discrete Poincar\'{e} symmetry, the quasi-energy-momentum is limited
within the Brillouin zone and transforms as Eq.~\eqref{eq:Ektransform}
or~\eqref{eq:Ebarkbartrans}. The single-particle Hilbert space is finite
if we carefully choose $\ket{p}$ so that it transforms into itself after
finite times of $\hat U(1,0,0)$ transformation.

The dispersion relation should be carefully designed so that the set $\left\{ \left(\bar{E}_{\bar k}, \bar{k}\right) \right\}$ is
invariant under the transformation~\eqref{eq:Ebarkbartrans} with $j=1$. Such a dispersion relation is
said to be Lorentz-invariant. We define $\tilde{k}=\bar k N$ which
is an integer ($ 0,1,\cdots,N-1$), and define $\tilde E= \bar E N$. Now Eq.~\eqref{eq:Ebarkbartrans} is rewritten as
\begin{equation}\label{eq:tildeEktrans}
\begin{split}
& \tilde E' \equiv \gamma(1)\tilde E+ \left(\gamma(1)^2-1\right) \tilde k  \ \  \ \left(\text{mod} \ N\right), \\
& \tilde k' \equiv \tilde E + \gamma(1) \tilde k   \ \  \ \left(\text{mod} \ N\right) ,
\end{split}
\end{equation}
where the left-hand and right-hand sides of the equal sign are congruent modulo $N$. According to
the second equation of~\eqref{eq:tildeEktrans}, $\tilde E$ must be an integer, and then $\tilde E$ can only take
$0, 1,\cdots,N-1$.

To obtain a basis, we assign an integer $ \tilde E_{\tilde k} \in [0,N)$ to each integer $ \tilde k \in [0,N)$
so that the $N$ integer pairs $\left(\tilde E_{\tilde k}, \tilde k\right) $ 
transform into each other under the transformation~\eqref{eq:tildeEktrans}.
The basis vectors $\ket{E_k,k}$ with $E_k = \displaystyle\frac{2\pi \tilde E_{\tilde k}}{T N}$
and $k = \displaystyle\frac{2\pi \tilde k}{a N}$ then meet both the orthonormal and invariant conditions.
We use them to generate a single-particle Hilbert space.

Now is a good time to revisit the time translations. For the quasi-energy chosen in this paper,
we can easily derive $\ket{m+N,n}=\ket{m,n}$ from Eq.~\eqref{eq:realdecomp} for arbitrary $m$ and $n$.
This means that the time translation of $N$ periods must be $\hat U(0,N,0)= 1$.
In other words, an arbitrary quantum state definitely evolves into itself after $N$ periods. Therefore,
we can choose a time interval of length $NT$ (say $[0,NT)$) and propose the periodic boundary
conditions along the $t$-axis, when discussing the quantum states or operators in real spacetime.
Two quantum states $\ket{m,n}$ and $\ket{m',n'}$ are equal to each other if $m \equiv m' \ \left(\text{mod} \ N\right)$
and $n \equiv n' \ \left(\text{mod} \ N\right)$. It is convenient to redefine the coordinate
transformation $P(j,m,n)$ by proposing an equivalence relation in the spacetime lattice $\mathcal{Y}$.
Two sites at $(mT,na)$ and $(m'T,n'a)$ are equivalent to each other if $m \equiv m' \ \left(\text{mod} \ N\right)$
and $n \equiv n' \ \left(\text{mod} \ N\right)$. With respect to this equivalence relation, the lattice $\mathcal{Y}$
becomes a square of edge $N$. The Lorentz boost $P(j,0,0)$ maps $(m,n)$ into $(m',n')$ with
\begin{equation}\label{eq:mnmpnptrans}
\begin{split}
& m' \equiv \gamma(j) m + \zeta(j)n  \ \ \left(\text{mod} \ N\right), \\
& n' \equiv \zeta(j) \left(\gamma(1)^2-1\right) m + \gamma(j) n \ \ \left(\text{mod} \ N\right).
\end{split}
\end{equation}
An interesting relation exists between the Lorentz transformations in real spacetime
and in quasi-energy-momentum space. If we exchange $\tilde k$ and $\tilde E$, the mappings
$\left(\tilde k, \tilde E \right) \to \left(\tilde k', \tilde E' \right)$ and $\left(m,n \right) \to 
\left(m',n' \right)$ indeed obey the same law (compare Eq.~\eqref{eq:mnmpnptrans}
with~\eqref{eq:Ebarkbartrans}).

Our method of constructing the Hilbert space
is not guaranteed to work for arbitrary $N$. Indeed, there exists $N$ for which no set of integer pairs
keeps invariant under the transformation~\eqref{eq:tildeEktrans}. In the case that such a set exists,
it may not be unique. Unfortunately, we do not find a simple criterion for judging the existence
of Lorentz-invariant dispersion relation for general $N$. But with the help of a computer,
we can enumerate these relations as $N$ is not big.

\subsection{The case $\gamma(1)=2$}

\begin{table}[b]
\renewcommand\arraystretch{2.0}
\begin{tabular}{| p{0.8 cm}<{\centering} | p{0.7 cm}<{\centering} | p{0.7 cm}<{\centering}
| p{0.7 cm}<{\centering} | p{0.7 cm}<{\centering} | p{0.7 cm}<{\centering} |
p{0.7 cm}<{\centering} | p{0.7 cm}<{\centering} | p{0.6 cm}<{\centering} |}
\hline
$j$ & 0 & 1 & 2 & 3 & 4 & 5 & 6 & $\cdots$ \\
\hline
$\gamma(j)$ & 1 & 2 & 7 & 26 & 97 & 362 & 1351 & $\cdots$ \\
\hline
$\zeta(j)$ & 0 & 1 & 4 & 15 & 56 & 209 & 780 & $\cdots$ \\
\hline
\end{tabular}
\caption{The first few elements of the sequences $\gamma(j)$ and $\zeta(j)$ as $\gamma(1)=2$.}\label{tab}
\end{table}
A few examples are helpful for explaining the above ideas. Let us consider the discrete Poincar\'{e} group $\mathcal{P}$
with $\gamma(1)=2$ (the smallest integer $\gamma(1)$).

The Lorentz matrix depends on $\gamma(j)$ and $\zeta(j)$.
Table~\ref{tab} lists the first few elements of the integer sequences $\gamma(j)$ and $\zeta(j)$ which increase
exponentially with $j$. Indeed, the ratios of $\gamma(j+1)$ to $\gamma(j)$ or $\zeta(j+1)$ to $\zeta(j)$
both converge to $\gamma(1)+\sqrt{\gamma(1)^2-1}\approx 3.732$ in the limit $j\to\infty$.
While the ratio of $\gamma(j)$ to $\zeta(j)$ converges to $\sqrt{\gamma(1)^2-1}\approx 1.732$.
On the other hand, the periods of space and time translations are $a$ and $T=\sqrt{3}a/c$, respectively.
The Lorentz matrix $L(j)$ transforms the spacetime point $\left(mT,na\right)$ into $\left(m'T,n'a\right)$ with
\begin{equation}\label{eq:gamma2Lorentz}
\left( \begin{array}{c} m'\\n' \end{array}\right) = \left( \begin{array}{cc} 2 & 1 \\ 3 & 2\end{array} \right)^j 
\left( \begin{array}{c} m \\n \end{array}\right).
\end{equation}

We consider a particle on a 1D lattice of $N$ sites under periodic boundary conditions.
The quasi-momentum of the particle can only take $N$ possible values which are $k={2\pi \tilde k}/{(a N)} $
with $\tilde k =0,1,\cdots,N-1$. The quasi-energy is $E_k = {2\pi \tilde E_{\tilde k}}/{(T N)}$,
where $ \tilde E_{\tilde k}$ is an integer between $0$ and $N-1$. The unitary operator $\hat U(1,0,0)$ transforms the
single-particle state $\ket{E_k, k}$ into $\ket{E'_{k'}, k'}$ with
\begin{equation}\label{eq:tildeEkg2}
\begin{split}
& \tilde E' \equiv 2 \tilde E+ 3 \tilde k  \ \  \ \left(\text{mod} \ N\right), \\
& \tilde k' \equiv \tilde E + 2 \tilde k   \ \  \ \left(\text{mod} \ N\right) .
\end{split}
\end{equation}
The dispersion relation $ \tilde E_{\tilde k}$ must keep invariant under the transformation~\eqref{eq:tildeEkg2}.

Let us first consider the smallest nontrivial dimension $N=2$. There exist two dispersion relations which are Lorentz-invariant.
We can choose either $\tilde E_{\tilde k}=0$ at $\tilde k=0$ and $\tilde E_{\tilde k}=1$ at $\tilde k=1$, or $\tilde E_{0}=1$
and $\tilde E_{1}=0$. The other two relations (i.e. $\tilde E_{0}=\tilde E_{1}=1$ and $\tilde E_{0}=\tilde E_{1}=0$)
do not keep invariant under the transformation~\eqref{eq:tildeEkg2}, therefore, they cannot be the dispersion relations
of a theory with discrete Lorentz symmetry.

\begin{table}[b]
\renewcommand\arraystretch{2.0}
\begin{tabular}{| p{0.6 cm}<{\centering} | p{0.6 cm}<{\centering} | p{0.6 cm}<{\centering} | 
p{0.6 cm}<{\centering} | p{0.6 cm}<{\centering} | p{0.6 cm}<{\centering} | 
p{0.6 cm}<{\centering} | p{0.6 cm}<{\centering} | p{0.6 cm}<{\centering} | p{0.6 cm}<{\centering} |}
\hline
2 & 3 & 6 & 11 & 13 & 18 & 22 \\
\hline
23 &  26  & 30 & 33 &  37 &  39 & 46  \\
\hline
47 & 54 & 59 & 61 & 66 & 69 & 71 \\
\hline
73 & 74 &  78 &  83 & 90 & 94 & 97 \\
\hline
\end{tabular}
\caption{The possible length ($N$) of a one-dimensional Bravais lattice that can
host a Lorentz-invariant quantum theory with $\gamma(1)=2$.}\label{tab:N}
\end{table}
For $N=3$, the unique Lorentz-invariant relation is $\tilde E_{0}= \tilde E_1 = \tilde E_2=0$,
i.e., the quasi-energy is zero at arbitrary quasi-momentum (a flat band). But for $N=4$ or $N=5$,
there exists no Lorentz-invariant $E_{k}$, hence, no quantum theory can be built.
Table~\ref{tab:N} lists all the possible $N$ ($N\leq 100$) for which the Lorentz-invariant $E_{k}$ exists.

\begin{figure}[tbp]
\includegraphics[width=0.9\linewidth]{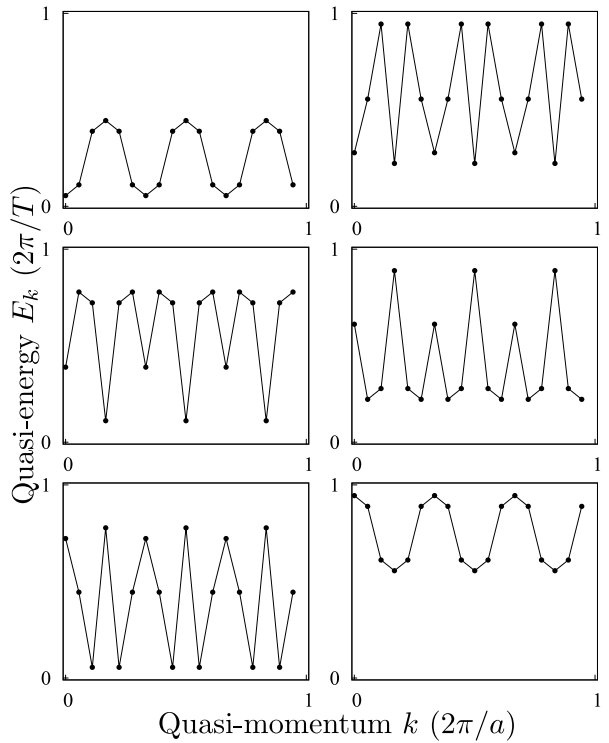}
\caption{The six Lorentz-invariant dispersion relations at $N=18$.}\label{fig:N18}
\end{figure}

\begin{figure}[tbp]
\includegraphics[width=0.9\linewidth]{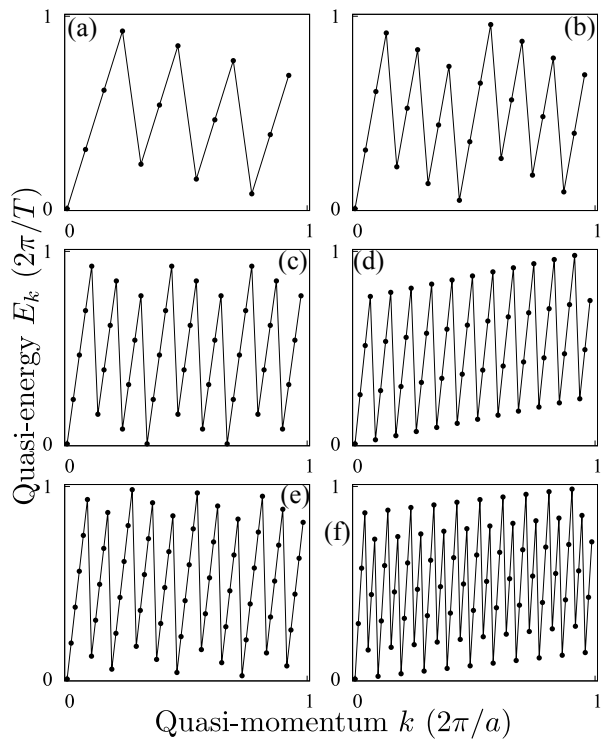}
\caption{The sawtooth Lorentz-invariant dispersion relations at (a)
$N=13$, (b) $N=23$, (c) $N=39$, (d) $N=47$, (e) $N=59$ and (f) $N=73$.}\label{fig:Nsawtooth}
\end{figure}
For most of $N$ in table~\ref{tab:N}, the Lorentz-invariant dispersion relations are not unique.
Fig.~\ref{fig:N18} plots the six different $E_k$ for $N=18$. They are symmetric with respect to $k=\pi/a$.
By using Eq.~\eqref{eq:tildeEkg2} one can prove that,
if $E_k$ is Lorentz-invariant, then both $E'_k = E_{2\pi/a-k}$ and
$E''_k= 2\pi/T-E_{k}$ are Lorentz-invariant (see Appendix~\ref{app:Eksymm} for the proof).
$E'_k$ and $E''_k$ are the mirror images of $E_k$
with respect to $k=\pi/a$ and $E=\pi/T$, respectively. In fig.~\ref{fig:N18},
the six functions are indeed the mirror images of each other with respect to $E=\pi/T$.

Fig.~\ref{fig:Nsawtooth} plots the dispersion relations
that have a sawtooth shape for different $N$. Such a sawtooth $E_k$ is found for many $N$
(e.g. $N=11,$ $13, $ $23, $ $39, $ $47, $ $59, $ $73,$ $83, \cdots$). The sawtooth $E_k$
is not symmetric with respect to $k=\pi/a$ or $E=\pi/T$, but it changes into itself after two successive reflections
to $k=\pi/a$ and $E=\pi/T$. Therefore, they always appear in pairs
($E_k$ and $E'_k = E_{2\pi/a-k}=2\pi/T-E_k$ are both Lorentz-invariant). Furthermore, the sawtooth $E_k$
all starts from $E_0=0$, as far as we know.

\section{Many-body theory \label{sec:Hamiltonian}}

\subsection{The field operators and unitary transformations}

The generalization from single-particle Hilbert space to many-body Hilbert space is straightforward.
In the many-body space, it is convenient to express the unitary transformations $\hat U(j,m,n)$ in terms of
the field operators.

We use $\hat a_k^\dag$ to denote the field operator that creates one particle of quasi-momentum $k$.
The single-particle state can then be expressed as
\begin{equation}\label{eq:akdefine}
\ket{E_k,k} = \hat a_k^\dag \ket{\text{vac}}
\end{equation}
with $\ket{\text{vac}}$ denoting the vacuum. The vacuum is a state without any particle,
which is defined by $\hat a_k \ket{\text{vac}}=0$ for arbitrary $k$. Here the particle can be either a fermion or
a boson. If it is a fermion, $\hat a_k^\dag$ satisfies the anticommutative relation, i.e.
$\left[ \hat a_k, \hat a_{k'}^\dag\right]_+ = \hat a_k \hat a_{k'} ^\dag + \hat a_{k'}^\dag \hat a_k = \delta_{k,k'} $.
If the particle is a boson, $\hat a_k^\dag$ satisfies the commutative relation, i.e.
$\left[\hat a_k, \hat a_{k'}^\dag\right] = \hat a_k \hat a_{k'} ^\dag - \hat a_{k'}^\dag \hat a_k = \delta_{k,k'} $.
The following results are independent of whether the commutative or anticommutative relations stand.
Therefore, we will not distinguish whether the particle is a fermion or boson any more.

The unitary transformation $\hat U(j,m,n)$ does not rotate the vacuum state, i.e.
$\hat U(j,m,n) \ket{\text{vac}}= \ket{\text{vac}}$. This is
guaranteed if the right-most field operator of $\hat U(j,m,n)$ is an annihilation operator.
Acting $\hat U(0,m,n)$ on both sides of Eq.~\eqref{eq:akdefine} and then using Eq.~\eqref{eq:eigenvalueUmn},
we obtain
\begin{equation}\label{eq:Uak}
\hat U(0,m,n) \hat a^\dag_k \hat U^{-1} (0,m,n) = e^{-inak+imTE_k} \hat a^\dag_k .
\end{equation}
Eq.~\eqref{eq:Uak} tells us how the field operator transforms under a coordinate translation.
For those who are familiar with quantum field theory, it is easy to find
\begin{equation}\label{eq:Umndef}
\hat U(0,m,n) = e^{-i n a \sum_k k \hat a^\dag_k \hat a_k + i mT \sum_k E_k \hat a^\dag_k \hat a_k }.
\end{equation}
Taking $m=0$ and $n=1$, we find the space-translational operator to be $\hat U(0,0,1)=
e^{-ia \sum_k k \hat a^\dag_k \hat a_k}$. It is clear that $\sum_k k \hat a^\dag_k \hat a_k$ must be
the quasi-momentum operator of the quantum field. 

Next we show how to express the unitary operator $\hat U(j,0,0)$ that corresponds
to a Lorentz boost. Since $\hat U(j,0,0)=\hat U^j(1,0,0)$, we focus on the expression of $\hat U(1,0,0)$.
By using the field operators, Eq.~\eqref{eq:Ujonp} can be reexpressed as
\begin{equation}\label{eq:U1ak}
\hat U(1,0,0) \hat a^\dag_k \hat U^{-1}(1,0,0) = \hat a^\dag_{k'},
\end{equation}
where the relation between $k'$ and $k$ has been given in Eq.~\eqref{eq:Ektransform} with $j=1$.
The transformation from $k=2\pi\tilde k/(aN)$ to $k'=2\pi\tilde k'/(aN)$ is also equivalently defined
in Eq.~\eqref{eq:tildeEktrans}. $k \to k'$ is indeed a permutation
of the set $\left\{ 0, \frac{2\pi}{a} \frac{1}{N},  \frac{2\pi}{a}\frac{2}{N}, \cdots, \frac{2\pi}{a}\frac{N-1}{N}\right\}$.
As is well known, a permutation of $N$ elements can be represented
by a $N$-by-$N$ matrix (so-called permutation matrix), denoted as $\sigma$, whose entry $\sigma_{k',k}$
is $1$ if $k$ maps to $k'$ but $0$ otherwise. $\sigma$ is a unitary matrix, therefore, it can always
be written as $\sigma= e^{-iK}$ with $K$ being a hermitian matrix. The Lorentz boost can then be
expressed as
\begin{equation}
\hat U(1,0,0) = e^{-i \sum_{k,k'} \hat a^\dag_{k'} K_{k',k} \hat a_k }.
\end{equation}

\subsection{The Floquet Hamiltonians}

Let us consider the time evolution operators.
Taking $m=-1$ and $n=0$ in Eq.~\eqref{eq:Umndef}, we find
$\hat U(0,-1,0)= e^{-iT \sum_k E_k \hat a^\dag_k \hat a_k}$. $\hat U(0,-1,0)$ is
the time-translational transformation, mapping the state seen by an observer $K$ to that seen by $K'$ (notice that
the Heisenberg picture is automatically chosen as we talk about the transformation between observers).
When the clock of $K$ reads $t=0$ (the reference time in the Heisenberg picture), the clock of $K'$ reads $t'=t-T=-T$.
Therefore, if we turn to the Schr\"{o}dinger picture, $\hat U(0,-1,0)$
must be the evolution operator over one period $T$. In the periodically-driven problems,
people often formally write the evolution operator as $e^{-iT \hat H_F}$
where $\hat H_F$ is the so-called Floquet Hamiltonian. Similarly, we introduce the Floquet Hamiltonian in our theory.
By comparing $e^{-iT \hat H_F}$ with $\hat U(0,-1,0)$, we find
\begin{equation}\label{eq:FH}
\hat H_F = \sum_k E_k \hat a^\dag_k \hat a_k.
\end{equation}
Notice that $\hat H_F$ is not defined in a unique way. One does not change the evolution operator
$\hat U(0,-1,0)$ by adding integer times of ${2\pi}/{T} $ to $E_k$.
In this paper, we choose $0\leq E_k < 2\pi/T$ to be in the Brillouin zone when defining
$\hat H_F$.

We emphasize that time is continuous. Only the time-translational symmetry is discrete.
We usually choose $t=0$ to be the reference time, hence, we can
obtain the quantum states in the Schr\"{o}dinger picture at $t= 0,
\pm T, \pm 2T, \cdots$ by acting $\hat U(0,m,0)$ on the state at $t=0$.
The quantum states not at these times are less interesting to us. They are not unambiguously defined
in the theory. For example, one cannot simply claim that the state at $t=T/3$ is $e^{-i \hat H_F T/3}$ acting on
the state at $t=0$. Indeed, the trajectory of the state from $t=0$ to $t=T$ can
be freely chosen, which is not constrained by the discrete Poincar\'{e} symmetry.

\begin{figure}[tbp]
\includegraphics[width=0.9\linewidth]{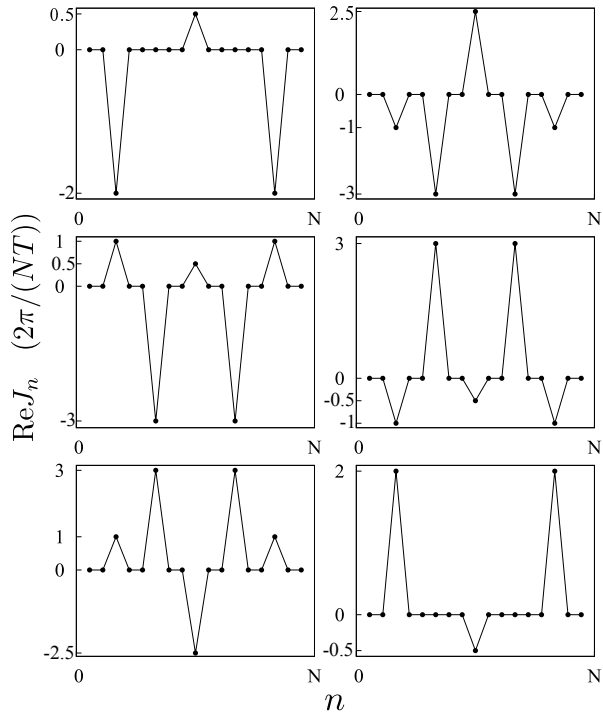}
\caption{The six coupling functions $J_n$ of the corresponding
Lorentz-invariant theories at $N=18$.}\label{fig:couplingN18}
\end{figure}

\begin{figure}[tbp]
\includegraphics[width=0.9\linewidth]{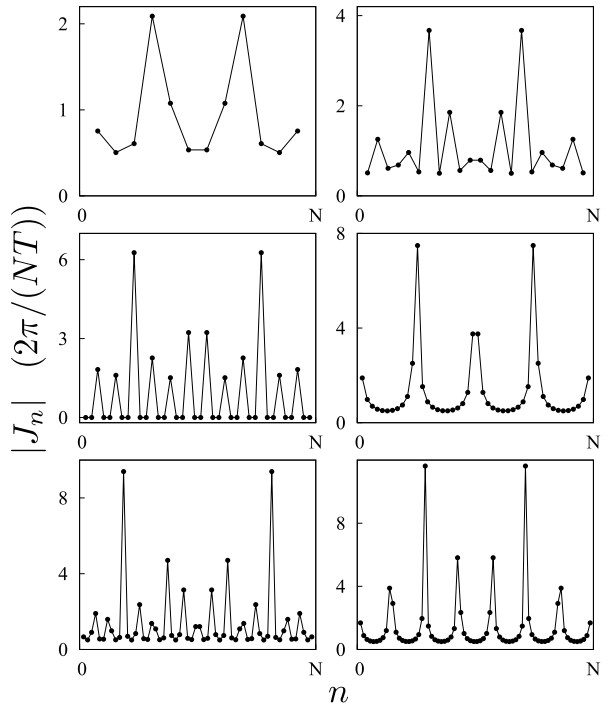}
\caption{The absolute coupling functions of a sawtooth dispersion relation at $N=13$,
$23$, $39$, $47$, $59$ and $73$, respectively.}\label{fig:couplingsawtooth}
\end{figure}

Sometimes, it is more convenient for designing an experiment setup if we express the Floquet Hamiltonian
in terms of the field operators in real space. To define the real-space field operators, we
notice that $\ket{m,n}$ with $m=0$ represents the quantum state of a particle located at site $n$.
We then define the creation operator at site $n$ according to
\begin{equation}
\ket{0,n} = \hat b^\dag_n \ket{\text{vac}}
\end{equation}
where $n=0,1,\cdots,N-1$. And $\hat b^\dag_n$ is either a fermionic operator satisfying $\left[\hat b_n ,\hat b^\dag_{n'}\right]_+
=\delta_{n,n'}$ or a bosonic operator satisfying $\left[\hat b_n ,\hat b^\dag_{n'}\right] =\delta_{n,n'}$.
By using Eq.~\eqref{eq:realdecomp} and~\eqref{eq:akdefine}, we connect $\hat b^\dag_n$ to
the field operators in quasi-momentum space:
\begin{equation}
\hat b^\dag_n = \frac{1}{\sqrt{N}} \sum_k e^{-inak} \hat a_k^\dag.
\end{equation}
The inverse transformation is $\hat a_k^\dag= \frac{1}{\sqrt{N}} \sum_n e^{inak}\hat b^\dag_n$.
We reexpress the Floquet Hamiltonian~\eqref{eq:FH} as
\begin{equation}\label{eq:HFJn}
\hat H_F = J_0 \sum_n  \hat b^\dag_n \hat b_n + \sum_{n > n'} \left(J_{n-n'} \hat b^\dag_n \hat b_{n'}
+ \text{h.c.} \right),
\end{equation}
where the first and second terms are the onsite potential and hopping between different sites, respectively.
$J_n$ with $n$ being an integer is the coupling function, defined as $J_n = \frac{1}{N} \sum_{k} E_k e^{iak n}$.
From the definition we immediately obtain $J_n=J^*_{N-n}$ where $J^*$ denotes the complex conjugate
of $J$. Since the quasi-energy $E_k$ can be increased
or decreased by arbitrary integer times of $2\pi/T$, the coupling function $J_n$ is also not uniquely defined.
Next we study $J_n$ as $E_k$ is in the Brillouin zone $B_f$.

The coupling function is determined by the dispersion relation, and then depends on $N$.
As a special example, we obtain $\hat H_F=0$ as $N=3$ ($E_k=0$). $\hat H_F=0$ means that the quantum state repeats itself
after one period of evolution.

Fig.~\ref{fig:couplingN18} plots $J_n$ at $n\neq 0$ for the six
different $E_k$ (see fig.~\ref{fig:N18}) as $N=18$, in which case $E_k=E_{2\pi/a-k}$ directly
indicates that $J_n$ is a real number. Interestingly, $J_n$ vanishes for most of $n$.
For example, in the top left panel, only $J_3=J_{15}=-4\pi/(NT)$ and $J_9=\pi/(NT)$ are nonzero.
This means that the particle can only hop between two sites at a distance of $3$ or $9$ from each other.
Notice that the distance between two sites being $15$ is equivalent to it being $3$ due to the periodic boundary condition.
In the bottom right panel, we see $J_3=J_{15}=4\pi/(NT)$ but $J_9=-\pi/(NT)$. Indeed, $E_k$
in the bottom right panel is the mirror image of that in the top left panel with respect to $E=2\pi/T$ (see fig.~\ref{fig:N18}).
It is easy to prove that, if $E'_k=2\pi/T-E_k$, then $J'_n=-J_n$.

Fig.~\ref{fig:couplingsawtooth} plots the absolute value of the coupling functions that correspond to the sawtooth
dispersion relations in fig.~\ref{fig:Nsawtooth}. Within the different $N$ in
fig.~\ref{fig:couplingsawtooth}, $N=39$ is special,
for which $J_n$ is zero at two thirds of $n$. But $J_n$ is nonzero for the other $N$ ($N=13$, $23$, $47$, $59$, $73$),
for which the dispersion relation satisfies $E_k\neq 0$ at $k\neq 0$ and $E_k = 2\pi/T-E_{2\pi/a-k}$.
From this property of $E_k$ we immediately derive $\text{Re} J_n = -\pi/(NT)$
where $\text{Re} J_n$ denotes the real part of $J_n$.
The absolute value $\left| J_n\right|$ displays a peak structure. Especially for $N=47$ and $N=73$,
we clearly see a few evenly-spaced sharp peaks. The coupling functions we find here are significantly different
from those in the other quantum many-body models. Not only the long-range hopping exists, but also the coupling
strength fluctuates with the distance.

\section{Green's functions \label{sec:green}}

In above, we have introduced the general approach to construct a quantum many-body theory
with discrete Poincar\'{e} symmetry on a 1D lattice of length $N$. The Floquet Hamiltonians for some specific $N$ have
been shown as examples. Next, we discuss the properties of such a theory. Especially, we
focus on the Green's functions in real spacetime.

\subsection{Field operators in real spacetime\label{sec:fieldreal}}

It is useful to show how the real-space field operators transform under a symmetry
transformation $\hat U(j,m,n)$. In the Heisenberg picture, the operator $\hat b^\dag_n(t)$ at
$t=mT$ ($\hat b_n^\dag (m)$ for short) is naturally defined as
$\hat b_n^\dag (m)=  e^{i mT\hat H_F} \hat b^\dag_n e^{-i mT\hat H_F}$.
And its relation with the single-particle state $\ket{m,n}$ is $\hat b_n^\dag (m) \ket{\text{vac}} = \ket{m,n}$.
From Eq.~\eqref{eq:realstatetranslation}, we immediately derive
\begin{equation}\label{eq:mnbn}
\begin{split}
\hat b_{n+n'}^\dag \left(m+m'\right)= \hat U(0,m',n') \ \hat b^\dag_n(m) \ \hat U^{-1} (0,m',n').
\end{split}
\end{equation}
In Eq.~\eqref{eq:mnbn}, if $n+n'$ is not in the range $[0,N)$, we need to replace it by
an integer congruent to $n+n'$ modulo $N$, because we choose the periodic boundary conditions along the $x$-axis.
Recalling $\hat U(0,N,0)=1$, we can also choose the periodic boundary conditions along the $t$-axis.
We then replace $m+m'$ by an integer congruent to it modulo $N$ if $m+m'$ is not in the range $[0,N)$.
In next, only the field operators $\hat b_n^\dag (m)$ with $0\leq n,m <N$ are considered.

There are a few ways to derive the transformation of $\hat b^\dag_n(m)$ under a Lorentz boost.
One can start from Eq.~\eqref{eq:U1ak} and notice that $\hat b^\dag_n$ is the Fourier transformation of $\hat a^\dag_k$.
By using the relation~\eqref{eq:pandpp}, we obtain
\begin{equation}
\hat U(j,0,0) \hat b^\dag_n(m) \hat U^{-1} (j,0,0) = \hat b^\dag_{n'}(m'),
\end{equation}
where $(m,n)\to(m',n')$ has been defined in Eq.~\eqref{eq:mnmpnptrans}.
The set of operators $\left\{ \hat b^\dag_{n'}(m') \right\}$ transforms under
$\hat U(j,m,n)$ in the same way as the spacetime
lattice $\left\{(m'T,n'a)\right\}$ equipped with a congruence relation (modulus $N$) transforms under $P(j,m,n)$.

\subsection{The Green's functions}

The Green's functions are the correlations between the field operators at two spacetime points.
Let us consider the retarded Green's function
\begin{equation}
G^r ( m_1,n_1,m_2,n_2) = -i \theta(m_1-m_2) \langle \left[ \hat b_{n_1}(m_1), \hat b^\dag_{n_2}(m_2) \right]_\mp \rangle,
\end{equation}
where the upper (lower) sign corresponds to bosons (fermions), respectively, the Heaviside function $\theta(\Delta m)$ is
$1$ if $\Delta m\geq 0$ but $0$ otherwise, and the brackets mean the expectation with
respect to some quantum state. $G^r$ is the correlation
between the field at position $x_1=n_1a$ at time $t_1=m_1T$ and that at position $x_2=n_2a$ at time $t_2=m_2T$.
In this paper, $\hat H_F$ has a quadratic form, therefore, $\left[ \hat b_{n_1}(m_1), \hat b^\dag_{n_2}(m_2) \right]_\mp$
is a number instead of an operator, and $G^r $ is then independent of the quantum state.

The discrete Poincar\'{e} symmetry manifests itself in the retarded Green's function.
The (anti)commutator transforms in the same way as the field operators, while the latter
transforms in the same way as the spacetime coordinates. We then have
\begin{equation}
\begin{split}
& \hat U(j,m,n) \left[ \hat b_{n_1}(m_1), \hat b^\dag_{n_2}(m_2) \right]_\mp \hat U^{-1}(j,m,n) \\ 
& =  \left[ \hat b_{n'_1}(m'_1), \hat b^\dag_{n'_2}(m'_2) \right]_\mp,
\end{split}
\end{equation}
where $(m_1,n_1)\to (m'_1,n'_1)$ or $(m_2,n_2)\to (m'_2,n'_2)$ is the coordinate transformation $P(j,m,n)$.
Since $G^r $ is independent of the quantum state, we obtain
\begin{equation}\label{eq:Grrelation}
\begin{split}
G^r ( m_1,n_1,m_2,n_2) = G^r ( m'_1,n'_1,m'_2,n'_2)
\end{split}
\end{equation}
for $m_1 \geq m_2$ and $m'_1 \geq m'_2$. Eq.~\eqref{eq:Grrelation} puts a set of
constraints on $G^r$ as we consider different $P(j,m,n)$.
The translational symmetry requires that $G^r$ should depend only upon the
difference of coordinates between two spacetime points. For simplicity, $G^r(m_1,n_1,m_2,n_2)$
is rewritten as $G^r(\Delta m, \Delta n)$ with $\Delta m=m_1-m_2$ and $\Delta n = n_1-n_2$.
According to the above discussions, we constrain the arguments $\Delta m$ and $\Delta n$ to be in the domain $[0,N-1)$.
More important, the Lorentz symmetry requires
\begin{equation}\label{eq:GrLorentzrelation}
\begin{split}
G^r ( \Delta m, \Delta n) = G^r ( \Delta m' , \Delta n'),
\end{split}
\end{equation}
where the map $( \Delta m, \Delta n)\to ( \Delta m' , \Delta n')$ is defined by Eq.~\eqref{eq:mnmpnptrans}.

\begin{figure}[tbp]
\includegraphics[width=0.9\linewidth]{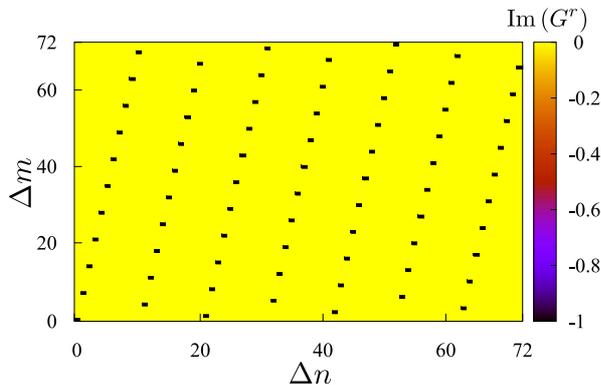}
\caption{(Color online) The imaginary part of $G^r$ as a function
of temporal difference $T\Delta m $ and spatial difference $a \Delta n $ for a Lorentz-invariant
theory on a lattice of length $N=73$. $G^r$ is nonzero only at a lattice of spacetime points (the black spots).}\label{fig:GrN73}
\end{figure}

The retarded Green's function evaluates
\begin{equation}
G^r(\Delta m, \Delta n)=-i \frac{1}{N} \sum_{\tilde k=0}^{N-1} e^{i2\pi
\left(\tilde k \Delta n - \tilde E_{\tilde k} \Delta m\right)/N}.
\end{equation}
The Green's function is uniquely defined. Choosing the quasi-energy-momentum
in different Brillouin zone may change the definition of $\hat H_F$, but it leaves $G^r$ unchanged.
Fig.~\ref{fig:GrN73} plots the imaginary part of $G^r(\Delta m,\Delta n)$ for $N=73$
and the dispersion relation fig.~\ref{fig:Nsawtooth}(f). This dispersion relation satisfies
$E_k = 2\pi/T - E_{2\pi/a-k}$, which directly indicates $G^r$ being purely imaginary.
We see that $G^r$ is zero almost everywhere except for $N$ specific $(\Delta m,\Delta n)$ where $G^r=-i$
(the black spots). The distribution of the black spots has the Lorentz symmetry, as required by Eq.~\eqref{eq:GrLorentzrelation}.
Their positions are indeed $\left(\Delta m,\Delta n\right)= \left( \tilde k, \tilde E_{\tilde k}\right)$.
This is not surprising, because we have shown below Eq.~\eqref{eq:mnmpnptrans} that
$\left( \tilde k, \tilde E_{\tilde k}\right)$ and $\left(\Delta m,\Delta n\right)$ transform in the same way
under the Lorentz boosts. Therefore, the set $\left\{\left( \tilde k, \tilde E_{\tilde k}\right)\right\}$
is Lorentz-invariant. The retarded Green's function tells us how a single particle propagate in the spacetime.
According to fig.~\ref{fig:GrN73}, the particle starts from $x=0$ at $t=0$, and jumps
from one site to the other after one entire period of evolution. The wave package keeps localized
as $t$ equals integer times of $T$.

\begin{figure}[tbp]
\includegraphics[width=0.9\linewidth]{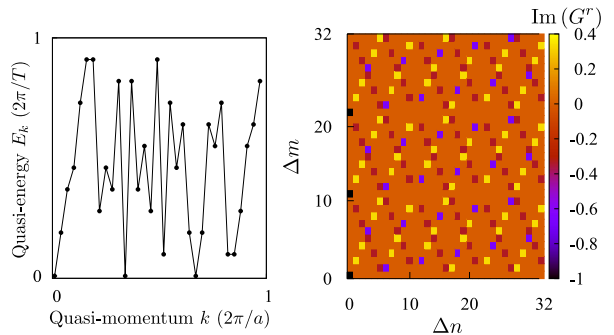}
\caption{(Color online) The left panel plots the dispersion relation of a Lorentz-invariant theory
on a lattice of length $N=33$. The right panel plots the corresponding Green's function $G^r(\Delta m,\Delta n)$.
$G^r$ is nonzero at some dispersed spots which are in black, purple, red or yellow.}\label{fig:GrN33}
\end{figure}

Fig.~\ref{fig:GrN33} the right panel gives another example of the Green's function. We choose $N=33$ and the
corresponding dispersion relation is plotted in the left panel. Again, $E_k = 2\pi/T - E_{2\pi/a-k}$
is satisfied, and then $G^r$ is purely imaginary. $G^r$ equals $-i$, $-2i/3$, $-i/3$ and $i/3$
at the spots colored black, purple, red and yellow, respectively, but zero everywhere else.
According to the Lorentz symmetry, the set of black spots must be Lorentz-invariant, so are the sets
of other colored spots. The difference between fig.~\ref{fig:GrN73} and~\ref{fig:GrN33} is that
$G^r$ of the latter is nonzero at multiple sites for a given $\Delta m$. This means that the wave
package spreads over a few sites after the evolution over one period. The black spots reappear at $\Delta m=11$
or $22$. The wave packages recombine into a single peak located at $\Delta n=0$ after every $11$ periods.

\section{Summary \label{sec:summary}}

In this paper, we review the discrete Poincar\'{e} group $\mathcal{P}$
and systematically develop the quantum theories with discrete Poincar\'{e} symmetry on
the one-dimensional lattice of constant $a$ and length $N$. $\mathcal{P}$ contains discrete Lorentz transformations
and discrete spacetime translations. The latter forms a Bravais lattice on the $t$-$x$ plane.
The elements of $\mathcal{P}$ are represented by unitary operators. The common eigenstate of translations
is defined to be a single-particle state with definite quasi-momentum $k$ and quasi-energy $E$. 
The discrete nature of translations confines $(E,k)$ in a single Brillouin zone - a primitive cell of
the reciprocal lattice. The multiplication law of group determines how $E$ and $k$ transform under a discrete Lorentz boost.
It is similar to the way that energy and momentum transform under a continuous Lorentz transformation.
But $(E,k)$ must be mapped back into the Brillouin zone after the transformation.
The dimension of the single-particle Hilbert space is $N$. A particle can be localized on
one of the $N$ sites in real space. The $N$ localized states are orthogonal to each other,
at the same time, the single-particle Hilbert space keeps invariant under a Lorentz boost.
These two conditions require that $k$ and $E$ be integer times of $2\pi/(Na)$ and $2\pi/(NT)$,
respectively, where $T$ is the period of time translation, and the dispersion relation $E_k$ be Lorentz-invariant.
For a given $N$, each Lorentz-invariant $E_k$ defines a quantum theory.
But for some $N$, no Lorentz-invariant $E_k$ can be found, thereafter, no Lorentz-invariant quantum lattice theory exists.
We enumerate the $28$ possible $N$ between $2$ and $100$ for $\gamma(1)=2$ where $\gamma(1)$
is the generator of the discrete Lorentz transformations. The corresponding $E_k$ displays reflection symmetry
with respect to $k=\pi/a$ or $E=\pi/T$. For some $N$, there exist a family of sawtooth dispersion relations in which
the pairs $(E_k,k)$ form a lattice in the Brillouin zone.

After we obtain the single-particle dispersion relation, it is straightforward to write down a
many-body theory of indistinguishable fermions or bosons. The field operators are defined both in
real space and in momentum space. The unitary operators of $\mathcal{P}$, i.e. the translations
and Lorentz boosts, are expressed in terms of the field operators. Especially, the discrete time-translational operator
is also the evolution operator over one period, from which we derive the Floquet Hamiltonian $\hat H_F$.
$\hat H_F$ depends on $E_k$, being quadratic before we consider the particle-particle interaction.
In real space, $\hat H_F$ includes the onsite
potential and hopping term. To preserve the Lorentz symmetry, there usually exists long-range hopping
and the coupling strength fluctuates with the distance between sites. Some specific models are studied. An interesting case
happens at $N=18$, in which the hopping is nonzero only between two sites at a distance of $3$ or $9$.
We calculate the retarded Green's function $G^r$ (the propagator), which is the correlation between two field operators
at different positions and times and depends only upon the difference $\Delta x$ and $\Delta t$.
Under the Lorentz boost, a field operator in real spacetime transforms in the same way as
a spacetime point. As a result, the discrete Lorentz symmetry manifests itself in $G^r$ which takes the same
value at a class of $(\Delta t,\Delta x)$ that map into each other under the Lorentz transformations.
In the model with a sawtooth dispersion relation, $G^r$ is zero almost everywhere except for
a lattice of $(\Delta t,\Delta x)$ which has the same shape as the lattice of $(E_k,k)$ in the Brillouin zone.
The particle propagates like a soliton even in the absence of any nonlinear mechanism.

The ultracold atomic gas in the periodically-driven optical lattice is a possible candidate for observing the discrete Poincar\'{e} symmetry.
The lattice constant $a$ and driving period $T$ can be chosen freely. Consequently, the
speed of constant becomes $c=\sqrt{\gamma(1)^2-1} \ a/T$, which may be much smaller than the speed of light.
As we have mentioned above, the value of $c$ has no influence on our theory and then can be chosen freely.
To realize the Lorentz invariance, the hopping between lattice sites needs to be carefully tuned.
The hopping as a function of distance is highly irregular in the models we have studied,
so that it is not very probable in experiments. But we have only explored a small fraction of Lorentz-invariant lattice models
by limiting ourselves on a short chain ($N< 100$) with a fixed generator ($\gamma(1)=2$).
Whether there exists a realizable model on a longer chain or with larger generators ($\gamma(1)=3, 4 ,\cdots$)
is an interesting open problem.
Finally, it is possible to add an onsite particle-particle interaction to $\hat H_F$ without breaking the
Lorentz symmetry. It has been known that a $\delta$-interaction does not break the continuous
Lorentz symmetry, because two particles located at the same spacetime point in one reference frame
are also at the same point in the other reference frames. Developing an interacting lattice theory with
discrete Poincar\'{e} symmetry is a challenging work in future.

\section*{Acknowledgement}
Pei Wang is supported by NSFC under Grant Nos.~11774315 and~11835011,
and by the Junior Associates program of the Abdus Salam International Center for Theoretical Physics.

\appendix

\section{Formulas about $\gamma(j)$ and $\zeta(j)$ \label{app:gamma}}

We list some useful formulas about $\gamma(j)$ and $\zeta(j)$ without proof.
$\zeta(j)$ can be redefined as
\begin{equation}
\begin{split}
\zeta(j)\sqrt{\gamma(1)^2-1} = & \frac{1}{2} \left(\gamma(1)+\sqrt{\gamma(1)^2-1}\right)^j \\
& - \frac{1}{2} \left(\gamma(1)-\sqrt{\gamma(1)^2-1}\right)^j.
\end{split}
\end{equation}
$\gamma(j)$ ($\zeta(j)$) is an even (odd) function of $j$, respectively. In other words, $\gamma(-j)=\gamma(j)$
but $\zeta(-j)=-\zeta(j)$. The joint iterative relations are
\begin{equation}
\begin{split}
& \zeta(j+1)=\gamma(1)\zeta(j)+\gamma(j), \\
& \zeta(j-1) = \gamma(1)\zeta(j)-\gamma(j).
\end{split}
\end{equation}
$\zeta(j)$ satisfies a more complicated relation:
\begin{equation}
\zeta(i+j+1)=\zeta(i+1)\zeta(j+1)-\zeta(i)\zeta(j),
\end{equation}
or in a matrix form:
\begin{equation}
\begin{split}
& \left( \begin{array}{cc} \zeta(i+1) & \zeta(i) \\ -\zeta(i) & -\zeta(i-1) \end{array} \right) 
\left( \begin{array}{cc} \zeta(j+1) & \zeta(j) \\ -\zeta(j) & -\zeta(j-1) \end{array} \right) \\
 & = \left( \begin{array}{cc} \zeta(i+j+1) & \zeta(i+j) \\ -\zeta(i+j) & -\zeta(i+j-1) \end{array} \right) .
\end{split}
\end{equation}

\section{Lorentz transformation of the quasi-energy-momentum \label{app:Ektran}}

Solving Eq.~\eqref{eq:pandpp} becomes easier if we use the tensor notation.
From Eq.~\eqref{eq:pandpp}, we immediately obtain
\begin{equation}\label{eq:2pidi}
\left(me +nf\right)_\mu p^\mu + l \ 2\pi = \left(m' e +n'f\right)_\mu p'^\mu ,
\end{equation}
where the lower and upper indices denote the components of a tensor, and $l$ is an arbitrary integer.
The Lorentz transformation in the tensor form is written as $L^\mu_{\ \nu}$. $\{ L^\mu_{\ \nu}\}$ is a collection
of $4$ numbers, which constitutes the Lorentz matrix~\eqref{eq:disLorentz}.
In Eq.~\eqref{eq:Pmultilaw}, the map from $(m,n)$ to $(m',n')$ is indeed defined by
the coordinate transformation of the Lorentz boost, reading
\begin{equation}
\left(m' e +n'f\right)^\mu =  L^\mu_{\ \nu} \left(me +nf\right)^\nu.
\end{equation}
Using $L^\mu_{\ \nu} L_\mu^{\ \nu'} = \delta^{\nu'}_\nu$, we obtain
\begin{equation}\label{eq:tenmn}
\left(me +nf\right)_\mu =   L^\nu_{\ \mu}\left(m' e +n'f\right)_\nu .
\end{equation}
Substituting Eq.~\eqref{eq:tenmn} into Eq.~\eqref{eq:2pidi}, we have
\begin{equation}
\left( p'^\mu-L^\mu_{\ \nu} p^\nu\right) \left(m' e +n'f\right)_\mu = l \ 2\pi.
\end{equation}
$\left(m' e +n'f\right)_\mu$ is the component of a lattice vector of $\mathcal{Y}$.
According to the definition of the reciprocal lattice $\mathcal{R}$, if the
components of a vector $r$ satisfy $r^\mu \left(m' e +n'f\right)_\mu$ being an integer times of $2\pi$,
$r$ must be a lattice vector of $\mathcal{R}$, i.e.
\begin{equation}\label{eq:pppmR}
\left( p'^\mu-L^\mu_{\ \nu} p^\nu\right) \in \mathcal{R}.
\end{equation}
Rewriting Eq.~\eqref{eq:pppmR} in terms of components, we obtain Eq.~\eqref{eq:Ektransform}.

\section{A theorem on the Lorentz-invariant dispersion relation\label{app:Eksymm}}

$E_k$ is a Lorentz-invariant dispersion relation, if $E_k$ and $k$ can be expressed as
$E_k={2\pi \tilde E_{\tilde k} }/{\left(TN\right)}$ and $k={2\pi \tilde k }/{\left(aN\right)}$, respectively,
where $\tilde E_{\tilde k}$ and $\tilde k$ are integers, and the set $\left\{ \left( \tilde E_{\tilde k}, \tilde k\right)\right\}$
keeps invariant under the map
\begin{equation}\label{app:tildeEktrans}
\begin{split}
 \tilde E' \equiv & \gamma(1)\tilde E+ \left(\gamma(1)^2-1\right) \tilde k  \ \  \ \left(\text{mod} \ N\right), \\
\tilde k' \equiv & \tilde E + \gamma(1) \tilde k   \ \  \ \left(\text{mod} \ N\right) .
\end{split}
\end{equation}
We will prove that $E'_k=E_{2\pi/a-k}$ and $E''_k = 2\pi/T- E_k$ are both
Lorentz-invariant if $E_k$ is Lorentz-invariant.
The relations $E'_k=E_{2\pi/a-k}$ and $E''_k = 2\pi/T- E_k$ can be translated into $\tilde E'_{\tilde k}=
\tilde E_{N-\tilde k}$ and $\tilde E''_{\tilde k} = N- \tilde E_{\tilde k}$, respectively. Therefore, we only need
to prove that the sets $\left\{ \left( \tilde E_{N-\tilde k}, \tilde k\right)\right\}$ and
$\left\{ \left( N- \tilde E_{\tilde k}, \tilde k\right)\right\}$ are both invariant under the map~\eqref{app:tildeEktrans}. 

For convenience in next, we use $\mathcal{L}$ to denote the map~\eqref{app:tildeEktrans}.
Suppose $\left( \tilde E, \tilde k\right)$ is an element of the set
$\left\{ \left( \tilde E_{\tilde k}, \tilde k\right)\right\}$ and $\mathcal{L}$ maps
$\left( \tilde E, \tilde k\right)$ into $\left( \tilde E', \tilde k'\right)$. Then
$\left( \tilde E', \tilde k'\right)$ must be an element of $\left\{ \left( \tilde E_{\tilde k}, \tilde k\right)\right\}$.
The converse is true. If $\left( \tilde E', \tilde k'\right)$ is an element of the set, then $\left( \tilde E, \tilde k\right)$
is also an element. This is because $\mathcal{L}$ is invertible. And the inverse map can be obtained by
solving Eq.~\eqref{app:tildeEktrans}, which reads
\begin{equation}\label{eq:invmap}
\begin{split}
\tilde E \equiv & \gamma(1)\tilde E'- \left(\gamma(1)^2-1\right) \tilde k'  \ \  \ \left(\text{mod} \ N\right), \\
\tilde k \equiv & -\tilde E' + \gamma(1) \tilde k'   \ \  \ \left(\text{mod} \ N\right) .
\end{split}
\end{equation}
From Eq.~\eqref{eq:invmap}, we immediately derive
\begin{equation}\label{app:invNm}
\begin{split}
 \tilde E \equiv & \gamma(1)\tilde E' + \left(\gamma(1)^2-1\right) \left(N-\tilde k' \right) \ \  \ \left(\text{mod} \ N\right), \\
 \left(N-\tilde k\right) \equiv & \tilde E' + \gamma(1) \left(N-\tilde k'\right)   \ \  \ \left(\text{mod} \ N\right) .
\end{split}
\end{equation}
If we compare Eq.~\eqref{app:tildeEktrans} with Eq.~\eqref{app:invNm}, the latter indeed tells us that $\mathcal{L}$
maps $\left(\tilde E', N-\tilde k'\right)$ into $\left(\tilde E, N-\tilde k\right)$. And $\mathcal{L}^{-1}$ then maps
$\left(\tilde E, N-\tilde k\right)$ into $\left(\tilde E', N-\tilde k'\right)$. Therefore, $\mathcal{L}$ or $\mathcal{L}^{-1}$
maps each element of the set $\left\{ \left( \tilde E_{N-\tilde k}, \tilde k\right)\right\}$ into another element in the same set.
In other words, $\left\{ \left( \tilde E_{N-\tilde k}, \tilde k\right)\right\}$ keeps invariant under $\mathcal{L}$.
Similarly, we can prove that $\left\{ \left( N- \tilde E_{\tilde k}, \tilde k\right)\right\}$ also keeps invariant under $\mathcal{L}$.

\end{document}